\documentclass[conference]{IEEEtran}
\IEEEoverridecommandlockouts
\usepackage{cite}
\usepackage{amsmath,amssymb,amsfonts}
\usepackage{algorithmic}
\usepackage{graphicx}
\usepackage{textcomp}
\usepackage{xcolor}
\usepackage[table]{xcolor}
\usepackage{multirow}
\usepackage{booktabs}
\def\BibTeX{{\rm B\kern-.05em{\sc i\kern-.025em b}\kern-.08em
    T\kern-.1667em\lower.7ex\hbox{E}\kern-.125emX}}

\usepackage{tikz}
\newcommand\copyrighttext{%
  \footnotesize © 2025 IEEE. Personal use of this material is permitted. Permission from IEEE must be obtained for all other uses, in any current or future media, including reprinting/republishing this material for advertising or promotional purposes, creating new collective works, for resale or redistribution to servers or lists, or reuse of any copyrighted component of this work in other works.}
\newcommand\copyrightnotice{%
\begin{tikzpicture}[remember picture,overlay]
  \node[anchor=south,yshift=10pt] at (current page.south) {\fbox{\parbox{\dimexpr\textwidth-\fboxsep-\fboxrule\relax}{\copyrighttext}}};
\end{tikzpicture}%
}
    
\begin{document}

\title{Evaluating Hardware Abstraction Layer Concepts for Software Defined Vehicles: Insights into Applicability and Effectiveness}

\author{
\IEEEauthorblockN{{Akshay Narla, Johannes Stümpfle, Souvik Saha, Nasser Jazdi and Michael Weyrich}}
\thanks{The authors would like to thank the German Federal Ministry of Education and Research (BMBF) (under Grant Number: 16MEE0472) and the Chips Joint Undertaking for the financial support under Grant Agreement No: 101139789 (HAL4SDV). The responsibility for the content of this publication lies with the authors.}
\IEEEauthorblockA{
\textit{Institute of Industrial Automation and Software Engineering (IAS)} \\
\textit{University of Stuttgart} \\
{Pfaffenwaldring 47, 70550 Stuttgart, Germany} \\
{E-Mail: \{akshay.narla, johannes.stuempfle, nasser.jazdi, michael.weyrich\}@ias.uni-stuttgart.de}}
}

\maketitle
\copyrightnotice

\begin{abstract}
The emergence of Software-Defined Vehicles (SDVs) represents a fundamental shift in automotive design, prioritizing software-centric architectures over traditional hardware-driven models. SDVs require modularity, interoperability, real-time processing, and over-the-air update capabilities throughout the vehicle lifecycle. However, current vehicle systems, characterized by tightly coupled software and hardware, struggle to meet these demands due to their complexity and heterogeneity. A critical first step toward enabling SDVs is the decoupling of software from hardware, which can be facilitated through a robust Hardware Abstraction Layer (HAL). While existing HALs offer hardware independence and standardized interfaces, their applicability and effectiveness in SDV contexts, characterized by multi-vendor ecosystems, safety-critical requirements, and cloud-edge usage, remain uncertain.

This paper systematically evaluates current automotive HALs and explores HAL mechanisms from non-automotive domains—including smartphones, networking, and industrial automation—to extract cross-domain insights relevant to SDV development. A criteria-driven evaluation framework is developed to assess HALs against SDV-specific needs. Findings reveal that while middleware-based HALs offer portability and modularity, hypervisor-based approaches better support safety, OTA readiness, and hardware efficiency. Limitations in both approaches are identified, prompting recommendations for a hybrid HAL design that integrates hypervisor isolation with middleware standardization. This paper contributes to the ongoing developments on automotive software architecture by offering insights into the applicability and effectiveness of current HAL strategies. It provides actionable guidance for designing flexible, scalable, and future-ready HALs to support SDVs across their lifecycle. The study supports industry-wide standardization efforts and aims to ensure long-term adaptability of automotive platforms in a connected and rapidly evolving mobility ecosystem.
\end{abstract}


\section{Introduction}

The automotive industry is transitioning from mechanically driven vehicles to SDVs, where software governs features, services, and value creation throughout the vehicle lifecycle. SDVs are expected to be continuously updateable, seamlessly connected, and integrated into the consumer’s digital ecosystem, enabling a connected mobility framework that spans in-vehicle systems, cloud services, and external infrastructures \cite{sdv_fev}. This transformation, driven by electrification, automation, shared mobility, and connected mobility, is projected to generate over \$650 billion for the automotive sector by 2030 \cite{bcgreport}. However, the complexity and heterogeneity of current vehicle hardware and software, coupled with fragmented interfaces and non-standardized data formats, hinder effective data exchange, real-time processing, and scalability required for future SDVs \cite{teixeira2025, federate}. This evolution requires a foundational shift in how software interacts with hardware in vehicles.

A critical barrier to realizing SDVs is the tightly coupled nature of software and hardware in existing vehicle architectures, which limits modularity and interoperability. To address this, a standardized HAL is essential to decouple software from hardware, enabling flexible, scalable, and interoperable software architectures. While HALs in the automotive domain provide hardware independence and standardized interfaces, their suitability for SDVs is uncertain due to challenges in managing multi-vendor ecosystems, real-time performance, safety, and integration with cloud and edge computing, among others \cite{federate}. Insights from HAL implementations in other domains, such as smartphones and the Internet, can offer solutions to some of these challenges.

This paper presents a comprehensive evaluation of existing HAL approaches within the automotive industry, while also identifying principles, mechanisms, and best practices from other domains to design an effective HAL tailored for SDVs. While prior studies have explored HALs in automotive contexts, they often focus on specific functionalities, such as device driver management, and lack a comprehensive analysis of SDV-specific requirements \cite{klüner2024}. A holistic evaluation of current HALs with respect to the emerging requirements of SDVs is missing. This work addresses this gap by an evaluation of HALs based on SDV needs and drawing cross-domain insights to provide design considerations and actionable recommendations for advancing automotive HAL.

The remainder of this paper is structured as follows: Section \ref{sec:rm} provides the research methodology followed while Section \ref{sec:hal} provides a brief overview of the HAL methods in non-automotive and automotive domains. Section \ref{sec:evaluation} presents the evaluation of current automotive HAL approaches with the set criteria and discusses its results in Section \ref{sec:discussion}. Finally, Section \ref{sec:conclusion} presents the conclusion and future works.

\section{Research Methodology}\label{sec:rm}
This study employs a systematic methodology to analyze HALs for SDVs, integrating a targeted review of selected publications and official technical documentation to develop evaluation criteria. The objective is to assess the suitability of existing HAL approaches for SDVs and provide architectural recommendations based on identified strengths and limitations.

A focused analysis of available resources was conducted to identify HAL implementations across non-automotive domains (e.g., smartphones, manufacturing, networking) and automotive contexts, using data from official documentation and select academic papers. This analysis established the current state-of-practice, highlighted existing gaps, and positioned the study within SDV research. Non-automotive HALs were examined to extract transferable best practices, while automotive HALs were evaluated for their alignment with SDV requirements. Observed challenges in middleware- and hypervisor-based approaches were synthesized to define the evaluation criteria. 

The MoSCoW method \cite{moscow} was adopted to prioritize criteria into Must-Have, Should-Have, and Could-Have categories, ensuring focus on critical SDV needs. Comparative analysis classified HALs into Standard API/Middleware-based and Hypervisor-based, enabling a structured evaluation framework. This methodology provides a robust foundation for evaluating HALs and informing SDV design, with findings synthesized in the discussion section to propose future directions.

\section{Hardware Abstraction Layer (HAL)}\label{sec:hal}
\begin{figure}[tbh]
    \vspace*{0.08in}
    \centering
    \includegraphics[width=0.65\linewidth]{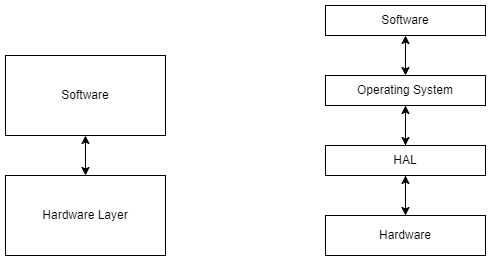}
    \caption{Idea of decoupling hardware and software with HAL}
    \label{fig:hal}
\end{figure}

HAL is a software layer that abstracts hardware-specific details of microcontrollers, processors, or other components, providing a standardized interface for operating systems (OS) and application software. By decoupling software from underlying hardware, as shown in Fig. \ref{fig:hal}, HALs enable portability, allowing applications to run across diverse hardware platforms with minimal modifications. This abstraction serves as a black box, shielding developers from low-level hardware complexities and enabling focus on application logic \cite{Yu_2022, hal_defn}. Key benefits of HALs include hardware independence, simplified development, enhanced maintainability, modularity, and support for performance optimization, making them essential for complex, heterogeneous systems \cite{Yu_2022, federate}.

In the context of SDVs, HAL plays a pivotal role in addressing the challenges of tightly coupled software-hardware architectures, multi-vendor hardware ecosystems, and the need for real-time performance, interoperability, and over-the-air updates \cite{bcgreport}. As SDVs evolve toward connected mobility ecosystems, requiring seamless integration with cloud services and external infrastructures, a robust HAL is critical to ensure scalability, safety, and standardization \cite{sdv_fev}. However, HAL implementations vary across domains, with different approaches to abstraction, each offering unique trade-offs in performance, flexibility, and complexity. This section provides an overview of HAL concepts, setting the foundation for a detailed analysis of non-automotive (Section \ref{subsec:nonauto}) and automotive (Section \ref{subsec:auto}) HAL approaches, which will help in providing insights to the design of a standardized HAL architecture tailored for SDVs.

\subsection{Non-Automotive HAL}\label{subsec:nonauto}
Non-automotive HALs in smartphones, manufacturing, networking, and cloud computing provide insights for modern HAL architectures, providing best practices for the decoupling of hardware and software in automotive. 

\textit{Android Project Treble} used an API-based HAL with Hardware Interface Definition Language (HIDL) to separate Android OS framework from vendor hardware code. Treble enabled rapid OTA updates and security patches, reducing deployment time. Its modular structure supports diverse hardware, improving update frequency \cite{treble}.

\textit{Asset Administration Shell (AAS)} is a core concept of the Industry 4.0 initiative that provides a digital representation of physical assets, designed to standardize and virtualize the interaction with industrial equipment. It abstracts hardware-specific interfaces through a middleware layer, which encapsulates data and functions of a device using an interoperable information model based on standards such as OPC UA \cite{aas}. 

\textit{Network Function Virtualization (NFV)} used hypervisors to virtualize network functions. This approach enabled faster upgrades, enhanced scalability, and consolidated resources, allowing multiple functions on a single server to reduce costs and improve system efficiency \cite{nfv}. 

\textit{Other Non-Automotive HALs:} VirtIO’s hypervisor-based HAL standardizes virtual device interfaces in cloud systems, aiding portability \cite{virtio}. Other HAL approaches analyzed include ARM CMSIS \cite{cmsis} for embedded portability, and Zephyr RTOS \cite{zephyr} for IoT connectivity.

\begin{figure}[tbh]
    \centering
    \includegraphics[width=0.94\linewidth]{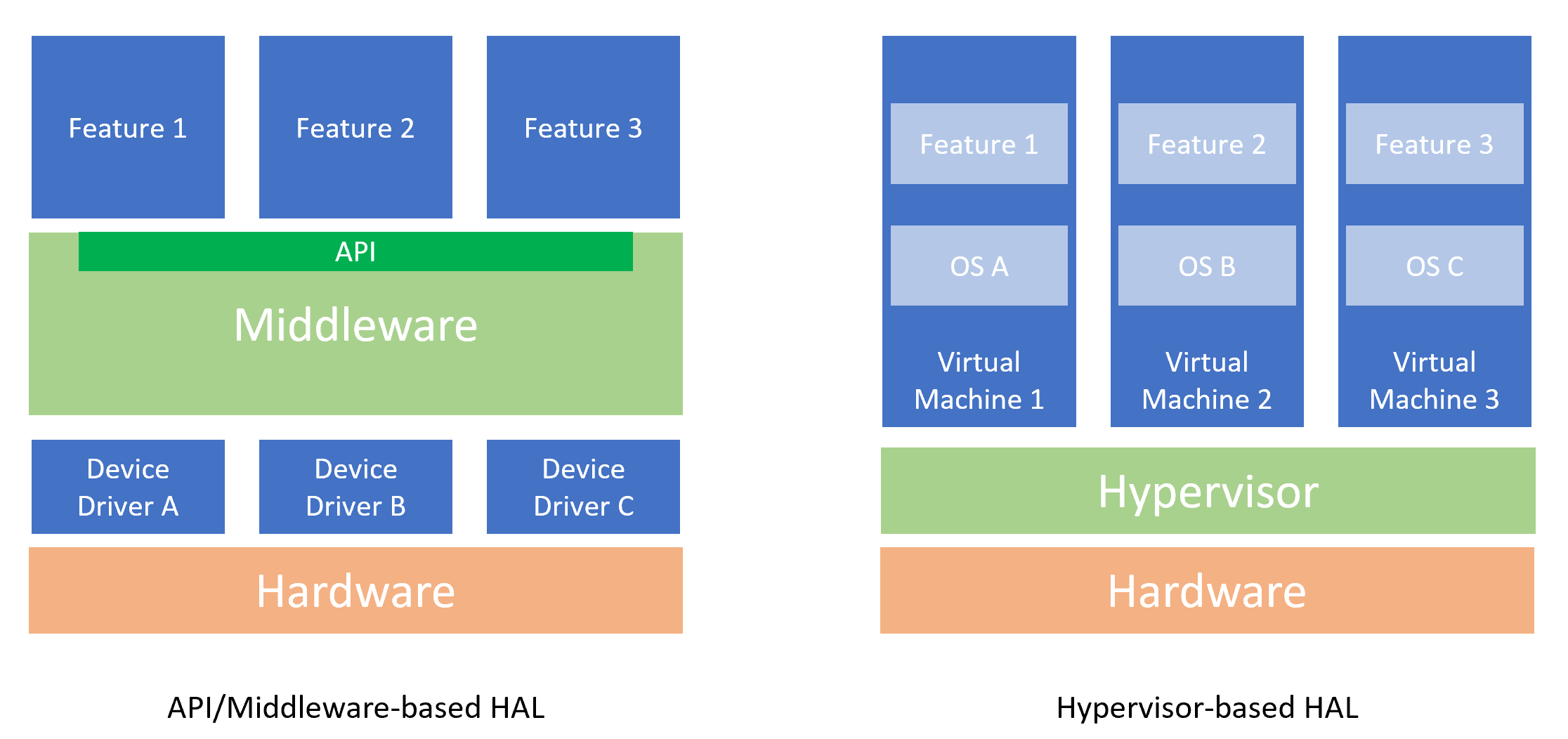}
    \caption{Main HAL Mechanisms}
    \label{fig:classify}
\end{figure}

\textit{\textbf{Observations:}} Based on our analysis, we can classify these HALs into two types based on their abstraction mechanisms, as shown in Fig. \ref{fig:classify}. Hypervisor-based HALs (NFV, VirtIO) embed abstraction within hypervisors, enabling resource sharing across virtual machines, which is advantageous for running multiple systems on a single processing unit. Standard API-based HALs (Android Treble, AAS, CMSIS, Zephyr) provide common APIs for hardware interaction, enhancing portability and development efficiency. However, applying these solutions to automotive contexts is not straightforward. In the following section, we discuss the various automotive HALs. 

\subsection{Automotive HAL}\label{subsec:auto}
Automotive HALs require stringent requirements in addition to other features of HAL. Aligning with non-automotive HAL classifications, we classify current automotive HAL methods or approaches also as Software Architecture/ Middleware-based and Hypervisor-based.

\subsubsection{\textbf{Middleware/API-based HAL}} Middleware/API-based HALs in the automotive domain aim to simplify development across diverse ECUs and vendors by abstracting hardware dependencies. These approaches typically use service-oriented architectures or standardized APIs to enable interoperability, OTA updates, and integration with cloud platforms. They support a flexible and modular software ecosystem—essential for realizing SDV functionalities—while attempting to meet safety and real-time constraints. \cite{klüner2024}

\textit{AUTOSAR Adaptive} uses a service-oriented architecture with middleware for high-performance applications. Its application layer supports event-driven communication via C++ APIs, the basic software layer provides system services like diagnostics and time synchronization, and the RTE layer interfaces components with hardware\cite{klüner2024}. 

\textit{Android Automotive} operates as a standalone HAL for in-vehicle infotainment, syncing with user Google accounts. It supports Play Store apps, enabling OTA updates and connections to external data like traffic reports. It reduces OEM development effort by providing a familiar interface \cite{aaos}. 

\textit{COVESA} uses a CommonAPI standard to develop distributed applications for connected vehicles. It standardizes vehicle data, aligning with W3C for a unified vehicle taxonomy. This is aimed at improving interoperability, enabling efficient data transmission to cloud services for vehicle-to-cloud systems \cite{covesa}. 

\textit{SOAFEE} is trying to combine hypervisors and middleware for cloud-native automotive systems, driven by real-world use cases. It aims to support safety-critical applications via ISO 26262 compliance, security through design principles and APIs, and meet real-time constraints by providing a scalable, secure platform for complex SDV applications\cite{soafee}. 

\textit{Apex.OS} is a safety-certified real-time operating system and development framework tailored for SDVs. Built as a fork of ROS 2, Apex.OS adds critical enhancements such as deterministic execution, real-time scheduling, and ISO 26262 ASIL D certification, making it suitable for safety-critical applications like autonomous driving \cite{becker2021}

\subsubsection{\textbf{Hypervisor-based HAL}}
Hypervisor-based HALs focus on consolidating heterogeneous systems by virtualizing hardware and enabling multiple OS environments to run in parallel on a single SoC. These approaches provide strong isolation between components, crucial for safety-critical systems in SDVs. They also allow legacy systems and new SDV software to coexist, which is valuable during the transition to fully software-defined architectures. \cite{otherhv}.

\textit{QNX Hypervisor} is a real-time hypervisor, runs applications with less than 2\% performance overhead. It supports virtual CPU models, shared graphics, and fast boot times with ISO 26262 compliance. Its versatile virtualization, reliability from QNX Neutrino RTOS, and system consolidation enable multiple OSs on a single SoC\cite{qnx}. 

\textit{EB Corbos}, an ASIL-B certified hypervisor, executes multiple guest OSs on a single CPU with isolated virtual machines. It supports cloud-based services, enhances safety and reliability, and scales for modern architectures. Its partnership-driven development ensures integration of industry innovations \cite{corbos}.

\begin{table*}[!t]
\centering
\caption{Evaluation Criteria for Automotive HAL}
\label{tab:evaluation_criteria}
\setlength{\tabcolsep}{5pt}
\begin{tabular}{@{\hspace{5pt}}l@{\hspace{5pt}}l@{\hspace{5pt}}p{\dimexpr0.65\textwidth-10pt}@{\hspace{5pt}}}
\toprule
\textbf{Category} & \textbf{Criterion} & \textbf{Definition} \\
\midrule

\multirow{6}{*}{Must-Have}& Handles Safety Requirements & Meets ISO 26262 with fail-safe mechanisms and fault detection. \\
    & OTA and Software Upgradability & Supports remote updates for functionality and security. \\
 & Security & Protects against cyberattacks, ensuring data integrity. \\
 & Low Latency & Minimizes delays for real-time applications like braking. \\
 & Compliance & Adheres to standards like AUTOSAR, MISRA for interoperability. \\
 & Multi-Platform Capability & Supports diverse hardware platforms for scalability. \\
 & Reduces Hardware Requirements & Optimizes hardware use to lower costs and power consumption. \\
\midrule

\multirow{3}{*}{Should-Have} & Modularity & Enables easy component replacement for maintainability. \\
 & Resource Utilization & Maximizes CPU/memory efficiency to avoid over-provisioning. \\
 & Scalability & Supports system expansion for new features or sensors. \\
\midrule

\multirow{4}{*}{Could-Have} & Developmental Ease & Provides tools to simplify development, reducing time/cost. \\
 & Cost Efficiency & Minimizes development/hardware costs for affordability. \\
 & Open Source & Supports community-driven standards, reducing licensing costs. \\
 & Flexibility & Adapts to new technologies or use cases for longevity. \\
\bottomrule
\end{tabular}
\end{table*}

\textit{Xen} is an open-source hypervisor using a microkernel to support fully and para-virtualized guests. Its Linux kernel provides I/O virtualization and reuses existing kernel drivers, enhancing efficiency for multi-platform virtualization \cite{xen}.

\textit{KVM}, a Linux-based hypervisor, integrates host and guest support in the Linux kernel and treats virtual machines as standard processes, utilizing kernel components like memory managers for efficient virtualization and scalability \cite{kvm}.

\textit{RTA-LWHVR} from ETAS, partitions ECUs into multiple VMs, enabling independent development and integration. It supports bare-metal applications, RTA-OS, or AUTOSAR stacks, preventing functional interference, protecting IP rights, and providing authentication to ensure software integrity \cite{rtalwh}.

\textit{\textbf{Observations:}} To determine the suitability of these HAL solutions for future SDVs, a structured evaluation is essential. While individual approaches offer strengths, such as modularity in middleware or isolation in hypervisors, none of the approaches single-handedly address the full spectrum of SDV requirements, which include safety, scalability, real-time performance, OTA capability, and multi-vendor support. Therefore, a comprehensive assessment based on well-defined criteria is necessary to identify gaps, overlaps, and opportunities for synergy. The following section presents this evaluation, aiming to establish if existing HAL approaches can provide a foundation for future SDVs or if they have some shortcomings.

\section{Evaluation Criteria for Automotive HALs}\label{sec:evaluation}

The previous section analyzed both non-automotive and automotive HAL approaches, highlighting their mechanisms and advantages. While non-automotive domains offer inspiration, such as modularity in Android Treble or virtualization in NFV, their direct applicability to automotive is constrained by safety-critical demands and real-time processing needs. Similarly, current automotive HAL implementations demonstrate promising directions but vary significantly in how well they address scalability, interoperability, and system consolidation.

To systematically compare these approaches and determine which are most suitable for future SDV architectures, a structured evaluation is necessary. This section presents a criteria-based evaluation framework, enabling objective assessment of the automotive HAL methods discussed earlier.

\subsection{Criteria Selection}

The MoSCoW method (Must-Have, Should-Have, Could-Have, Won’t-Have) \cite{moscow} was chosen to prioritize criteria due to its reliability in distinguishing critical requirements from desirable enhancements, ensuring focus on SDV necessities like safety and real-time performance. Criteria were selected based on challenges identified in automotive HAL implementations, including real-time execution, communication, security, and resource management for Middleware, and resource distribution, VM sprawl, and performance monitoring for Hypervisor-based methods \cite{klüner2024, virtchall}. These challenges, coupled with system requirements for safety, scalability, and interoperability, informed the selection of criteria to address shortcomings and align with industry standards (e.g., ISO 26262). 

The 7 Must‑Have criteria capture requirements that an SDV cannot operate without; an HAL that fails any of them is flagged as unsuitable for SDVs in its current state, even if it scores well elsewhere. 3 Should‑Have criteria provide a competitive advantage, and 4 Could‑Have criteria add supplementary value to the HAL. The complete set of criteria and their MoSCoW classification is presented in \ref{tab:evaluation_criteria}. 

\begin{figure*}[t]
    \vspace*{0.08in}
    \centering
    \includegraphics[width=0.88\linewidth]{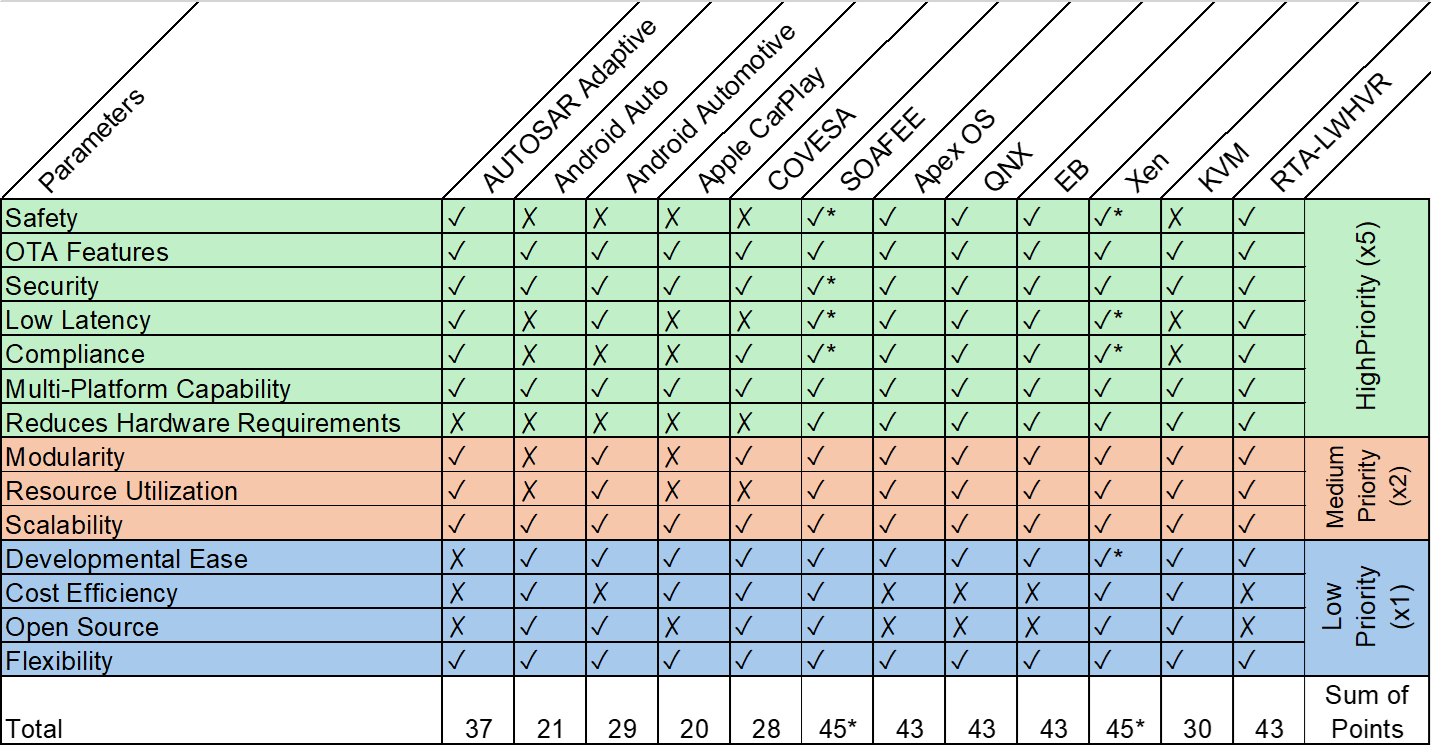}
    \caption{Evaluation of Automotive HALs against the criteria set in Table \ref{tab:evaluation_criteria}}
    \label{fig:eval_hal}
\end{figure*}

\subsection{Criteria Evaluation}
The scoring system assigns 5 points for Must-Have, 2 for Should-Have, and 1 for Could-Have criteria, making a possible maximum of 45 points for a HAL method, enabling quantitative comparison of automotive HALs. If a criterion is not met (marked as \textit{X}), the HAL simply receives 0 points for that entry, which already penalises the aggregate result, so an additional penalty‑point scheme was not considered. Asterisks (*) denote features that are currently under development but are expected to comply in future releases. Each HAL is assessed against these criteria defined in Table \ref{tab:evaluation_criteria} to identify strengths, gaps, and alignment with SDV requirements. The weighted scores are summed across these criteria, providing the final points for the HAL approach. A higher number is desirable as it indicates that the approach satisfies a large number of SDV requirements, provided all the must-have criteria are satisfied. This structured approach ensures a comprehensive analysis, enabling a transparent comparison of HALs and guiding the development of optimized SDV architectures.

\section{Discussion} \label{sec:discussion}


With the MoSCoW-weighted framework (Table \ref{tab:evaluation_criteria}) in place, the results in Fig. \ref{fig:eval_hal} translate published results such as peer-reviewed articles, certifications, benchmarks, and design documents into quantitative performance indicators. An \textit{X} cell always reflects missing public proof. Therefore, any HAL that omits even one must-have is flagged non-compliant based on its current state, as also indicated in Section \ref{sec:evaluation}.

Hypervisor-based methods generally outperformed middleware-based approaches in must-have criteria such as safety, OTA features, security, and hardware efficiency. Solutions like QNX, EB Corbos, and RTA-LWHVR, designed for safety-critical automotive systems as proprietary software, excel in safety compliance, aligning closely with ISO 26262 standards, satisfying all the must-have criteria \cite{qnx,corbos,otherhv}. However, open-source hypervisors like Xen and KVM lag in safety, low latency, and compliance, causing a 15-point reduction, making them incompatible in their current state. But they show promise, as indicated by Xen’s ongoing efforts toward automotive certification \cite{xen}.

Middleware-based HALs show more varied performance with AUTOSAR Adaptive, the industry standard, only meeting 6 of the 7 Must-Have criteria, with its own architecture conceding significant CPU and memory demand, in addition to being cost-inefficient, on high-performance ECUs, giving it 37 points out of 45 \cite{klüner2024, federate}. Apex.OS, a ROS 2 derivative certified to ASIL-D, closes the remaining high-priority gaps and rises to 43, its only deduction coming from proprietary tooling costs\cite{becker2021}.

By contrast, Android Automotive earns ticks for seamless OTA updates, vehicle-bus isolation, and hardware abstraction, yet the absence of any ISO 26262 artefact or bounded-latency benchmark leaves half of the Must-Have column empty, limiting the score to 29 \cite{aaos, treble}. Android Auto and Apple CarPlay, primarily designed for infotainment mirroring, scored poorly due to a lack of access to vehicle networks and limited real-time guarantees \cite{ivi}.

COVESA provides an abstraction layer but is mainly used for standardizing vehicle signals, as a complement to other HAL methods. Hence, it scores very less as its main purpose is different \cite{covesa}. Finally, SOAFEE demonstrates significant potential by combining hypervisor-level isolation with cloud-native middleware architectures. Although SOAFEE is still under development (denoted with *), it is expected to achieve full 45 points upon completion, particularly in safety, security, and compliance as mentioned in their documentation \cite{soafee}. This positions SOAFEE as a promising open-source alternative once development stabilizes.

Overall, the evaluation highlights that no single existing HAL fully meets all SDV requirements yet. Hypervisors lead in safety and scalability but struggle in cost and openness, while middleware offers modularity and developmental ease but often compromises on resource utilization and hardware isolation.

\subsection{Future implications for Automotive HAL}
The insights gained reveal that SDVs must manage increasing system complexity, strict real-time requirements, and evolving safety standards. This demands a rethinking of HAL architectures \cite{sdv_fev}. Our evaluation suggests a hybrid HAL model that leverages hypervisor-based isolation for critical systems and middleware-based standardization for flexibility and reuse. Hypervisors can consolidate hardware by running multiple OSs on fewer ECUs, improving efficiency and maintaining safety boundaries. Meanwhile, middleware such as Apex.OS or AUTOSAR Adaptive can streamline communication across applications, manage legacy integration, and facilitate OTA updates.

This separation ensures that failures in infotainment or non-critical systems do not compromise core vehicle functionality. Retaining familiar platforms like Android Automotive for UX while enforcing strict safety through hypervisors or certified middleware can accelerate development without sacrificing integrity. Moving forward, efforts should focus on reducing middleware resource overhead and advancing safety certification for open-source hypervisors like Xen to ensure affordability and openness in SDV deployments.

\subsection{Learnings from Non-Automotive HAL}
Non-automotive HALs, mentioned in Section \ref{subsec:nonauto}, offer mechanisms to address some of the SDV HAL challenges while meeting some of the criteria set in Table \ref{tab:evaluation_criteria} like scalability and low latency. Project Treble’s HIDL-based API model offers a promising reference for accelerating OTA updates in SDVs. AAS middleware highlights the potential of standardized communication across heterogeneous ECUs. NFV’s hypervisor virtualization suggests resource-sharing strategies to optimize hardware use. VirtIO’s standardized interfaces inspire approaches for enhancing platform portability. CMSIS points to methods for simplifying embedded development, while Zephyr RTOS showcases lightweight connectivity concepts valuable for real-time sensor data handling. Together, these cross-domain innovations provide a rich set of tools and paradigms that automotive HAL designers can adopt or adapt, making future SDV architectures more scalable, modular, and responsive to fast-evolving requirements.

\section{Conclusion} \label{sec:conclusion}
This paper provided insights into the applicability and effectiveness of existing HAL mechanisms for advancing SDV architectures. Through a criteria-driven evaluation of non-automotive HALs and automotive HALs, the study identified trade-offs between modularity, safety, real-time performance, and scalability. As defined in Table \ref{tab:evaluation_criteria} and visualized in Fig. \ref{fig:eval_hal}, middleware-based approaches offer high portability and abstraction but often fall short on safety and resource efficiency, whereas hypervisor-based solutions demonstrate superior isolation, scalability, and multi-domain security. Building on these findings, the study suggests a hybrid HAL architecture that integrates the strengths of hypervisors and middleware. This hybrid model will be capable of addressing key SDV challenges, such as managing system complexity, ensuring functional safety, enabling OTA updates, and supporting interoperability across diverse vendor ecosystems.

The main contributions of this work are as follows:
\begin{itemize}
    \item Developed a comprehensive evaluation framework for automotive HALs, enabling objective comparisons.
    \item Identified advantages of non-automotive HALs, such as interoperability and portability, for SDV design.
    \item Suggests a hybrid HAL architecture integrating hypervisor isolation with middleware standardization.
\end{itemize}

Future work will focus on implementing and validating the proposed hybrid HAL architecture within real SDV prototypes. Integration with the Eclipse SDV toolchain is planned to ensure open-source development and facilitate adoption by the broader automotive ecosystem. Overall, this work advances the understanding of HAL applicability and effectiveness, offers actionable recommendations for next-generation SDV design, and contributes to the standardization efforts required for a connected and continuously evolving automotive ecosystem.

\bibliographystyle{IEEEtran}
\bibliography{bib}

@techreport{bcgreport,
    author = "A. Koster et al.",
    title = "Rewriting the Rules of Software-Defined Vehicles",
    institution = "Boston Consulting Group: https://www.bcg.com/publications/2023/
rewriting-rules-of-software-defined-vehicles",
    year = "2023"
}

@InProceedings{sdv_fev,
author="Ke{\ss}ler, G{\"u}nter
and Sieben, Dominik
and Bhange, Anand
and B{\"o}rner, Elmar",
title="The Software Defined Vehicle -- Technical and Organizational Challenges and Opportunities",
booktitle="23. Internationales Stuttgarter Symposium",
year="2023",
isbn="978-3-658-42048-2"
}

@misc{teixeira2025,
      title={Deterministic and Reliable Software-Defined Vehicles: key building blocks, challenges, and vision}, 
      author={Pedro Veloso Teixeira and Duarte Raposo and Rui Lopes and Susana Sargento},
      year={2025},
      url={https://arxiv.org/abs/2407.17287}, 
}

@article{federate,
    author = "Peltonen, Ella and et al.",
    title = "Software-Defined Vehicle Support and Coordination Project",
    journal = "Federate Project Forecast Report",
    year = "2024"
}

@misc{klüner2024,
      title={Modern Middlewares for Automated Vehicles: A Tutorial}, 
      author={David Philipp Klüner and Marius Molz and Alexandru Kampmann and Stefan Kowalewski and Bassam Alrifaee},
      year={2024},
      url={https://arxiv.org/abs/2412.07817}, 
}

@techreport{hal_defn,
    author = "Peng, Hao and Dömer, R",
    title = "Towards a Unified Hardware Abstraction Layer	Architecture for Embedded Systems",
    institution = "CECS, UC Irvine",
    year = "2012"
}

@inproceedings{Yu_2022, 
    title={The Digital Foundation Platform -A Multi-Layered SOA Architecture for Intelligent Connected Vehicle Operating System},
   DOI={10.4271/2022-01-0107   },
   booktitle={SAE Technical Paper Series},
   author={Yu, David and Xiao, Andy},
   year={2022}}

@misc{moscow,
      title={Choosing a Suitable Requirement Prioritization Method: A Survey}, 
      author={Esraa Alhenawi and Shatha Awawdeh and Ruba Abu Khurma and Maribel García-Arenas and Pedro A. Castillo and Amjad Hudaib},
      year={2024},
      url={https://arxiv.org/abs/2402.13149}, 
}

@misc{treble,
    title = "Android Open Source Project",
    author = "Android Developers",
    url = "https://source.android.com/",
}

@misc{nfv,
    title = "What is NFV?",
    author = "Redhat",
    url = "https://www.redhat.com/en/topics/virtualization/what-is-nfv",
}

@misc{cmsis,
    title = "CMSIS",
    author = "ARM",
    url = "https://arm-software.github.io/CMSIS\_6/latest/General/index.html",
}

@INPROCEEDINGS{virtio,
  author={Ampelikiotis, Timos and Rigo, Alvise and Raho, Daniel},
  booktitle={ICAI}, 
  title={Technical overview and performance evaluation of Virtio-loopback}, 
  year={2024},
}

@ARTICLE{aas,
  author={de Oliveira, Vitor Furlan and Matiolli, Guilherme and Júnior, Cláudio José Bordin and Gaspar, Ricardo and Lins, Romulo Gonçalves},
  journal={IEEE Transactions on Intelligent Vehicles}, 
  title={Digital Twin and Cyber-Physical System Integration in Commercial Vehicles: Latest Concepts, Challenges and Opportunities}, 
  year={2024},
}

@misc{zephyr,
    title = "Zephyr Project Documentation",
    author = "Zephyr Project Members",
    url = "https://docs.zephyrproject.org/latest/"
}

@inproceedings{aaos,
  author       = {Abdul Moiz and
                  Manar H. Alalfi},
  title        = {A survey of security vulnerabilities in Android automotive apps},
  booktitle    = {Proceedings of EnCyCriS, Pittsburgh, USA},
year="2022"
}

@INPROCEEDINGS{covesa,
  author={Aust, Stefan},
  booktitle={2022 25th International Symposium on Wireless Personal Multimedia Communications (WPMC)}, 
  title={Vehicle API and Service Catalog for Next Generation Mobility}, 
  year={2022},
  }

@misc{soafee,
    title = "SOAFEE Architecutre Documentation",
    author = "SOAFEE Project Members",
    url = "https://architecture.docs.soafee.io/en/latest/"
}

@misc{qnx,
    title = "QNX Hypervisor",
    author = "Blackberry",
    url = "https://blackberry.qnx.com/en/products/foundation-software/qnx-hypervisor"
}

@misc{corbos,
    title = "Solid foundation for automotive software platform",
    author = "Elektrobit",
    url = "https://www.elektrobit.com/products/ecu/eb-corbos/"
}

@INPROCEEDINGS{otherhv,
  author={Zhang, Zhengjun and Liu, Yanqiang and Chen, Jiangtao and Qi, Zhengwei and Zhang, Yifeng and Liu, Huai},
  booktitle={27th International Conference on Parallel and Distributed Systems}, 
  title={Performance Analysis of Open-Source Hypervisors for Automotive Systems}, 
  year={2021},
}

@misc{virtchall,
    title = "7 Common Virtualization Challenges – And How to Overcome Them",
    author = "GDH",
    url = "https://gdhinc.com/7-common-virtualization-challenges-and-how-to-overcome-them/"
}

@InProceedings{becker2021,
author="Becker, Jan
and Sagar, Mehul
and Pangercic, Dejan",
editor="Bertram, Torsten",
title="A Safety-Certified Vehicle OS to Enable Software-Defined Vehicles",
booktitle="Automatisiertes Fahren 2021",
year="2021",
}

@misc{xen,
    title = "Xen Hypervisor Documentation",
    author = "Xen project members",
    url = "https://wiki.xenproject.org/wiki/Xen\_Project\_Release\_Features"
}

@misc{kvm,
      title={Virtualization and Microservice Architecture for Software-Defined Vehicles: An Evaluation and Exploration}, 
      author={Long Wen and Markus Rickert and Fengjunjie Pan and Jianjie Lin and Yu Zhang and Tobias Betz and Alois Knoll},
      year={2024}, 
}

@ARTICLE{rtalwh,
  author={Lozano, Santiago and Lugo, Tamara and Carretero, Jesús},
  journal={IEEE Access}, 
  title={A Comprehensive Survey on the Use of Hypervisors in Safety-Critical Systems}, 
  year={2023},
}

@misc{ivi,
      title={GM Says It's Ditching Apple CarPlay and Android Auto for Your Safety}, 
      author={Scott Evans},
      year={2023}, 
}

\end{document}